\documentclass{ws-p8-50x6-00}
\makeatletter\providecommand{\@ptsize}{0}\makeatother
\usepackage{graphicx,amsmath,amssymb,feynmp,emp}
\empaddtoTeX{\usepackage{amsmath,amssymb}}
\newtheorem{theorem}{Theorem}
\begin{document}
\begin{fmffile}{groves4sitgespics}
\fmfset{arrow_ang}{10}
\fmfset{curly_len}{2mm}
\fmfset{wiggly_len}{3mm}
\begin{empfile} 

%%%%%%%%%%%%%%%%%%%%%%%%%%%%%%%%%%%%%%%%%%%%%%%%%%%%%%%%%%%%%%%%%%%%%%%%
\title{%
  \makebox[\textwidth]{{}\hfill\texttt{IKDA 99/17}}\\
  \hfil\\
  Forests~\&~Groves:
  Minimal Gauge~Invariant~Classes
  of Tree~Diagrams in Gauge~Theories}
\author{Edward Boos}
\address{%
  Institute of Nuclear Physics,
  Moscow State University, 119899, Moscow, Russia\\
  E-mail: boos@theory.npi.msu.su}
\author{\underline{Thorsten Ohl}}
\address{%
  Darmstadt University of Technology,
  Schlo\ss gartenstr. 9, D-64289 Darmstadt, Germany\\
  E-mail: ohl@hep.tu-darmstadt.de}
\maketitle
%%%%%%%%%%%%%%%%%%%%%%%%%%%%%%%%%%%%%%%%%%%%%%%%%%%%%%%%%%%%%%%%%%%%%%%%

\abstracts{%
  We describe the explicit construction of \emph{groves}, the smallest
  gauge invariant classes of tree Feynman diagrams in gauge theories.}

\section{Introduction}

Calculations of cross section with many-particle final states remain
challenging despite all technical advances and it is crucial to be
able to concentrate on the important parts of the scattering amplitude
for the phenomena under consideration.  In gauge theories, however, it
is impossible to naively select a few signal diagrams and to ignore
the rest.  The same subtle cancellations among the diagrams in a gauge
invariant subset that lead to the celebrated good high energy behavior
of gauge theories come back to haunt us if we accidentally select a
subset of diagrams that is not gauge invariant.  Results of such a
calculation have \emph{no} predictive power, because they depend on
unphysical parameters introduced during the gauge fixing of the
Lagrangian.

The subsets of Feynman diagrams selected for any calculation must
therefore form a \emph{gauge invariant subset}, i.\,e.~ they must
satisfy the Ward- and Slavnov-Taylor-identities by themselves to
ensure the cancellation of contributions from unphysical degrees of
freedom.  Since not all diagrams in a gauge invariant subset have the
same pole structure, a selection based on ``signal'' or ``background''
will \emph{not} suffice.

In abelian gauge theories, such as QED, the classification of gauge
invariant subsets is straightforward and can be summarized by the
requirement of inserting any additional photon into \emph{all}
connected charged propagators.  This situation is similar for gauge
theories with simple gauge groups, the difference being that the gauge
bosons are carrying charge themselves.  For non-simple gauge groups
like the Standard Model~(SM), the classification of gauge invariant
subsets is more involved.

We present the explicit construction of
\emph{groves}\cite{Boos/Ohl:1999:groves}, the smallest
gauge invariant classes of tree Feynman diagrams in gauge theories.
This construction is applicable to gauge groups with any
number of factors, which can even be mixed, as in the~SM.
Furthermore, it does not require a summation over complete multiplets
and can therefore be used in flavor physics when members of weak
isospin doublets are detected separately.

In unbroken gauge theories, the permutation symmetry of external gauge
quantum numbers can be used to subdivide the scattering amplitude
corresponding to a grove further into gauge invariant sub-amplitudes.
In this decomposition, each Feynman diagram contributes to more than
one sub-amplitude.  It is not yet known how to perform a similar
decomposition systematically in the~SM, because the entanglement of
gauge couplings and gauge boson masses complicates the structure of
the amplitudes.

%%%%%%%%%%%%%%%%%%%%%%%%%%%%%%%%%%%%%%%%%%%%%%%%%%%%%%%%%%%%%%%%%%%%%%%%
\section{Gauge Cancellations}

Even if the general arguments are well known, the intricate nature of
the cancellations of unphysical contributions complicates the
development of systematic calculational procedures.  Nevertheless,
phenomenology needs numerically well behaved matrix elements, which
should be as compact as possible and the development of an automated
procedure remains a formidable challenge.  The selection of gauge
invariant subsets of Feynman diagrams described here is one necessary
step towards this goal.  One particular problem is that the Ward
identities relate diagrams with different pole structure.  For
example, in~$q\bar q\to gg$
\begin{equation*}
  \parbox{26\unitlength}{%
    \begin{fmfgraph}(25,12)
      \fmfleft{i1,i2}
      \fmfright{o4,o3}
      \fmf{fermion}{i1,v1}
      \fmf{fermion,tension=0.5}{v1,v4}
      \fmfdot{v1,v4}
      \fmf{fermion}{v4,i2}
      \fmf{gluon}{o3,v4}
      \fmf{gluon}{o4,v1}
    \end{fmfgraph}} + 
  \parbox{26\unitlength}{%
    \begin{fmfgraph}(25,12)
      \fmfleft{i1,i2}
      \fmfright{o3,o4}
      \fmf{fermion}{i1,v1}
      \fmf{fermion,tension=0.5}{v1,v4}
      \fmfdot{v1,v4}
      \fmf{fermion}{v4,i2}
      \fmf{phantom}{o4,v4}
      \fmf{phantom}{o3,v1}
      \fmffreeze
      \fmf{gluon}{o4,v1}
      \fmf{gluon,rubout}{o3,v4}
    \end{fmfgraph}} + 
  \parbox{26\unitlength}{%
    \begin{fmfgraph}(25,12)
      \fmfleft{i1,i2}
      \fmfright{o3,o4}
      \fmf{fermion}{i1,v1}
      \fmf{fermion}{v1,i2}
      \fmf{gluon}{v5,v1}
      \fmfdot{v1,v5}
      \fmf{gluon}{o4,v5}
      \fmf{gluon}{o3,v5}
    \end{fmfgraph}}
\end{equation*}
the numerator factors cancel parts of denominators and the Ward identity
is satisfied only by the sum and not by individual diagrams.

Therefore, the physical amplitudes are determined by an intricate web
of kinematical structure and gauge structure.  Nevertheless, the
identification of partial sums of Feynman diagrams that are gauge
invariant by themselves is of great practical importance.  It is more
economical to spent the time improving Monte Carlo statistics for the
important pieces of the amplitude than to calculate the complete
amplitude all the time. This requires the identification of the
smallest gauge invariant part of the amplitude that contains the
important pieces. Secondly, different parts of the amplitude are
characterized by different scales for radiative corrections, but
different scales can only be used if the parts correspond to
separately gauge invariant pieces.

%%%%%%%%%%%%%%%%%%%%%%%%%%%%%%%%%%%%%%%%%%%%%%%%%%%%%%%%%%%%%%%%%%%%%%%%
\section{Flavor Selection Rules}
\label{sec:FlavorSelectionRules}

One method of identifying gauge invariant subsets is to identify
subsets that contribute to a particular final
state\cite{Bardin/etal:1994:4f-classification,Boos/Ohl:1997:gg4f}.
These subsets are not minimal, but they show that selection rules of
flavor symmetries that commute with the gauge group are useful tools.
The simplest example is Bhabha scattering: the $s$-
and $t$-channel diagrams have to be gauge-invariant separately,
because both~$e^+e^-\to \mu^+\mu^-$ and~$e^+\mu^-\to e^+\mu^-$ are
physical processes with gauge invariant amplitudes.
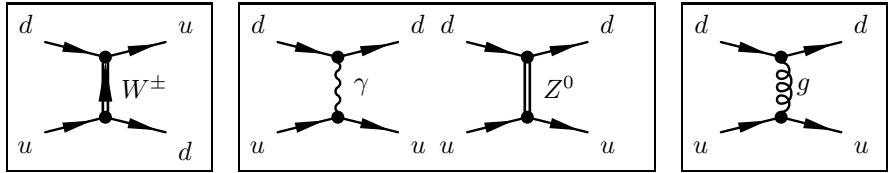
\begin{figure}
  \begin{center}
    \fbox{%
      \parbox{25\unitlength}{%
        \fmfframe(2,4)(2,4){%
          \begin{fmfgraph*}(20,12)
            \fmfleft{u1,d1}
            \fmflabel{$u$}{u1}
            \fmflabel{$d$}{d1}
            \fmfright{d2,u2}
            \fmflabel{$u$}{u2}
            \fmflabel{$d$}{d2}
            \fmf{dbl_plain_arrow,tension=0.5,label=$W^\pm$}{uv,dv}
            \fmf{fermion}{u1,uv,d2}
            \fmf{fermion}{d1,dv,u2}
            \fmfdot{uv,dv}
          \end{fmfgraph*}}}}\quad
    \fbox{%
      \parbox{53\unitlength}{%
        \fmfframe(2,4)(2,4){%
          \begin{fmfgraph*}(20,12)
            \fmfleft{u1,d1}
            \fmflabel{$u$}{u1}
            \fmflabel{$d$}{d1}
            \fmfright{u2,d2}
            \fmflabel{$u$}{u2}
            \fmflabel{$d$}{d2}
            \fmf{photon,tension=0.5,label=$\gamma$}{uv,dv}
            \fmf{fermion}{u1,uv,u2}
            \fmf{fermion}{d1,dv,d2}
            \fmfdot{uv,dv}
          \end{fmfgraph*}}
        \fmfframe(2,4)(2,4){%
          \begin{fmfgraph*}(20,12)
            \fmfleft{u1,d1}
            \fmflabel{$u$}{u1}
            \fmflabel{$d$}{d1}
            \fmfright{u2,d2}
            \fmflabel{$u$}{u2}
            \fmflabel{$d$}{d2}
            \fmf{dbl_plain,tension=0.5,label=$Z^0$}{uv,dv}
            \fmf{fermion}{u1,uv,u2}
            \fmf{fermion}{d1,dv,d2}
            \fmfdot{uv,dv}
          \end{fmfgraph*}}}}\quad
    \fbox{%
      \parbox{25\unitlength}{%
        \fmfframe(2,4)(2,4){%
          \begin{fmfgraph*}(20,12)
            \fmfleft{u1,d1}
            \fmflabel{$u$}{u1}
            \fmflabel{$d$}{d1}
            \fmfright{u2,d2}
            \fmflabel{$u$}{u2}
            \fmflabel{$d$}{d2}
            \fmf{gluon,tension=0.5,label=$g$}{uv,dv}
            \fmf{fermion}{u1,uv,u2}
            \fmf{fermion}{d1,dv,d2}
            \fmfdot{uv,dv}
          \end{fmfgraph*}}}}
  \end{center}
  \caption{\label{fig:ud2ud}%
    Gauge invariant subsets of Feynman diagrams in $ud\to ud$ scattering.}
\end{figure}

A less trivial example are the three separately gauge
invariant sets in $ud\to ud$~scattering shown in
figure~\ref{fig:ud2ud}.  The separate gauge invariance of the gluon
exchange diagram is obvious because the strong coupling can be
changed without violating gauge invariance.  The charged
current diagram is separately gauge-invariant, because we may assume
that the CKM mixing matrix is diagonal and then the charged current
diagram is absent in $us\to us$~scattering, which is related to
$ud\to ud$ by a horizontal symmetry that commutes with the gauge
group.

%%%%%%%%%%%%%%%%%%%%%%%%%%%%%%%%%%%%%%%%%%%%%%%%%%%%%%%%%%%%%%%%%%%%%%%%
\section{Forests}
\label{sec:forests}
To develop tools for taming the combinatorics, we
start from unflavored scalar $\phi^3$- and $\phi^4$-theory.  Since
there are no selection rules, the diagrams~$S_1$, $S_2$, and~$S_3$ in
\begin{equation}
\label{eq:flips}
  T_4 = \{t_4^{S,1},t_4^{S,2},t_4^{S,3},t_4^{S,4}\} = \left\{
   \parbox{16\unitlength}{%
     \begin{fmfgraph}(15,10)
       \fmfleft{g1,g2}
       \fmfright{g1',g2'}
       \fmf{plain}{g1,v,g2}
       \fmf{plain}{g1',v',g2'}
       \fmf{plain}{v,v'}
       \fmfdot{v,v'}
     \end{fmfgraph}},
   \parbox{16\unitlength}{%
     \begin{fmfgraph}(15,10)
       \fmfleft{g1,g2}
       \fmfright{g1',g2'}
       \fmf{plain}{g1,v1,g1'}
       \fmf{plain}{g2,v2,g2'}
       \fmf{plain}{v1,v2}
       \fmfdot{v1,v2}
     \end{fmfgraph}},
   \parbox{16\unitlength}{%
     \begin{fmfgraph}(15,10)
       \fmfleft{g1,g2}
       \fmfright{g1',g2'}
       \fmf{phantom}{g1,v1,g1'}
       \fmf{phantom}{g2,v2,g2'}
       \fmf{plain}{v1,v2}
       \fmfdot{v1,v2}
       \fmffreeze
       \fmf{plain}{g1,v1,g2'}
       \fmf{plain,rubout}{g2,v2,g1'}
     \end{fmfgraph}},
   \parbox{16\unitlength}{%
     \begin{fmfgraph}(15,10)
       \fmfleft{g1,g2}
       \fmfright{g1',g2'}
       \fmf{plain}{g1,v,g2}
       \fmf{plain}{g1',v,g2'}
       \fmfdot{v,v}
     \end{fmfgraph}}
    \right\}.
\end{equation}
must have the same coupling strength to ensure crossing invariance.
If there are additional symmetries, as in the case of gauge theories,
the coupling of~$t_4^{S,4}$ will also be fixed relative
to~$t_4^{S,1}$, $t_4^{S,2}$, and~$t_4^{S,3}$.

We call each exchange $t\leftrightarrow t'$ of two members of the
set~$T_4$ of all tree graphs with four external particles an
\emph{elementary flip}.  The elementary flips define a trivial
relation~$t\circ t'$ on~$T_4$, which is true, if and only if~$t$
and~$t'$ are related by a flip.  The relation~$\circ$ is trivial
on~$T_4$ because \emph{all} pairs are related by an elementary flip.
However, the elementary flips in~$T_4$ induce \emph{flips} in~$T_5$
(the set of all tree diagrams with five external particles) if they
are applied to an arbitrary four particle subdiagram
\begin{subequations}
\begin{equation}
   \parbox{19\unitlength}{%
     \begin{fmfgraph}(18,12)
       \fmfleft{g1,g2}
       \fmfright{g1',g2',g3'}
       \fmf{plain}{g1,v,g2}
       \fmf{plain,tension=0.5}{g1',v'}
       \fmf{plain}{v',v''}
       \fmf{dashes,tension=0.5}{v'',g2'}
       \fmf{dashes,tension=0.5}{v'',g3'}
       \fmf{plain}{v,v'}
       \fmfdot{v,v',v''}
     \end{fmfgraph}}
  \Longrightarrow
  \left\{
   \parbox{19\unitlength}{%
     \begin{fmfgraph}(18,12)
       \fmfleft{g1,g2}
       \fmfright{g1',g2',g3'}
       \fmf{plain}{g1,v1}
       \fmf{plain,tension=0.5}{v1,g1'}
       \fmf{plain}{g2,v2,v''}
       \fmf{dashes,tension=0.5}{v'',g2'}
       \fmf{dashes,tension=0.5}{v'',g3'}
       \fmf{plain}{v1,v2}
       \fmfdot{v1,v2,v''}
     \end{fmfgraph}},
   \parbox{19\unitlength}{%
     \begin{fmfgraph}(18,12)
       \fmfleft{g1,g2}
       \fmfright{g1',g2',g3'}
       \fmf{phantom}{g1,v1}
       \fmf{phantom,tension=0.5}{v1,g1'}
       \fmf{phantom}{g2,v2,v''}
       \fmf{phantom,tension=0.5}{v'',g2'}
       \fmf{dashes,tension=0.5}{v'',g3'}
       \fmf{plain}{v1,v2}
       \fmfdot{v1,v2,v''}
       \fmffreeze
       \fmf{dashes}{v'',g2'}
       \fmf{plain}{g1,v1,v''}
       \fmf{plain,rubout}{g2,v2,g1'}
     \end{fmfgraph}},
   \parbox{19\unitlength}{%
     \begin{fmfgraph}(18,12)
       \fmfleft{g1,g2}
       \fmfright{g1',g2',g3'}
       \fmf{plain}{g1,v,g2}
       \fmf{plain,tension=0.5}{g1',v}
       \fmf{plain}{v,v''}
       \fmf{dashes,tension=0.5}{v'',g2'}
       \fmf{dashes,tension=0.5}{v'',g3'}
       \fmfdot{v,v''}
     \end{fmfgraph}}
    \right\}
\end{equation}
Obviously, there is more than one element of~$T_4$ embedded in a
particular $t\in T_5$ and the same diagram is member of other
quartets, e.\,g.
\begin{equation}
   \parbox{19\unitlength}{%
     \begin{fmfgraph}(18,12)
       \fmfleft{g1,g2}
       \fmfright{g1',g2',g3'}
       \fmf{dashes}{g1,v,g2}
       \fmf{plain,tension=0.5}{g1',v'}
       \fmf{plain}{v',v''}
       \fmf{plain,tension=0.5}{v'',g2'}
       \fmf{plain,tension=0.5}{v'',g3'}
       \fmf{plain}{v,v'}
       \fmfdot{v,v',v''}
     \end{fmfgraph}}
  \Longrightarrow
  \left\{
   \parbox{19\unitlength}{%
     \begin{fmfgraph}(18,12)
       \fmfleft{g1,g2}
       \fmfright{g1',g2',g3'}
       \fmf{dashes}{g1,v,g2}
       \fmf{plain}{v,v'}
       \fmf{plain,tension=0.5}{v',g3'}
       \fmf{plain,tension=0.5}{g1',v'',g2'}
       \fmf{plain}{v',v''}
       \fmfdot{v,v',v''}
     \end{fmfgraph}},
   \parbox{19\unitlength}{%
     \begin{fmfgraph}(18,12)
       \fmfleft{g1,g2}
       \fmfright{g1',g2',g3'}
       \fmf{dashes}{g1,v,g2}
       \fmf{phantom}{v,v'}
       \fmf{phantom,tension=0.5}{v',g3'}
       \fmf{phantom,tension=0.5}{g1',v'',g2'}
       \fmf{plain}{v',v''}
       \fmffreeze
       \fmf{plain}{v,v',g2'}
       \fmf{plain,rubout}{g1',v'',g3'}
       \fmfdot{v,v',v''}
     \end{fmfgraph}},
   \parbox{19\unitlength}{%
     \begin{fmfgraph}(18,12)
       \fmfleft{g1,g2}
       \fmfright{g1',g2',g3'}
       \fmf{dashes}{g1,v,g2}
       \fmf{plain}{v,v'}
       \fmf{plain,tension=0.5}{v',g3'}
       \fmf{plain,tension=0.5}{g1',v',g2'}
       \fmfdot{v,v'}
     \end{fmfgraph}}
    \right\}
\end{equation}
\end{subequations}
\begin{figure}
  \begin{center}
    \hspace*{10mm}
    \begin{emp}(40,32)
      pair t_i[], t_o[];
      for i := -10 upto 10:
        t_o[i] := (0,min(w,h)) rotated -36i xscaled (w/h) shifted (w,h)/2;
      endfor
      t_i[0] := .5[t_o[-3],t_o[3]];
      t_i[1] := .3[t_o[5],t_o[2]];
      t_i[-1] := .3[t_o[-5],t_o[-2]];
      t_i[2] := .4[t_o[0],t_o[3]];
      t_i[-2] := .4[t_o[0],t_o[-3]];
      pickup pencircle scaled 0.8pt;
      draw for i := 0 upto 9: t_o[i]-- endfor
          t_o[0]--t_i[2]--t_o[3]--t_i[0]--t_o[-2]
        --t_i[-1]--t_o[-5]--t_i[1]
        --t_i[-1]--t_o[-1]--t_i[2]--t_o[4]
        --t_o[-4]--t_i[-2]--t_o[1]--t_i[1]
        --t_o[2]--t_i[0]--t_o[-3]--t_i[-2]--cycle;
      pickup pencircle scaled 3pt;
      for i := -4 upto 5: drawdot t_o[i]; endfor
      for i := -2 upto 2: drawdot t_i[i]; endfor
      label.llft (btex $\scriptstyle (23)(45) $ etex, t_o[-4]);
      label.llft (btex $\scriptstyle 2(3(45)) $ etex, t_o[-3]);
      label.ulft (btex $\scriptstyle 2((34)5) $ etex, t_o[-2]);
      label.ulft (btex $\scriptstyle (25)(34) $ etex, t_o[-1]);
      label.top  (btex $\scriptstyle ((25)4)3 $ etex, t_o[0]);
      label.urt  (btex $\scriptstyle ((24)5)3 $ etex, t_o[1]);
      label.urt  (btex $\scriptstyle (24)(35) $ etex, t_o[2]);
      label.lrt  (btex $\scriptstyle (2(35))4 $ etex, t_o[3]);
      label.lrt  (btex $\scriptstyle ((23)5)4 $ etex, t_o[4]);
      label.bot  (btex $\scriptstyle ((23)4)5 $ etex, t_o[5]);
      label.ulft (btex $\scriptstyle (2(45))3 $ etex, t_i[-2]);
      label lft  (btex $\scriptstyle (2(34))5 $ etex, t_i[-1]);
      label.bot  (btex $\scriptstyle 2((35)4) $ etex, t_i[0]);
      label.rt   (btex $\scriptstyle ((24)3)5 $ etex, t_i[1]);
      label.urt  (btex $\scriptstyle ((25)3)4 $ etex, t_i[2]);
      setbounds currentpicture to (0,-5)--(w,-5)--(w,h)--(0,h)--cycle;
    \end{emp}
    \hspace*{15mm}
    \includegraphics[width=.4\textwidth]{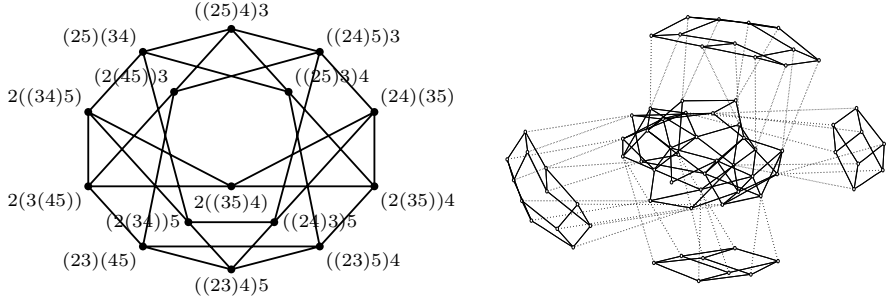}
  \end{center}
  \caption{\label{fig:5phi3/mix71}%
    Left: the forest~$F_5$ of the 15 five-point tree diagrams in unflavored
    $\phi^3$-theory. The diagrams are specified by fixing vertex~1 and
    using parentheses to denote the order in which lines are joined at
    vertices.
    Right: the forest of size~71 for the
    process~$\gamma\gamma\to u\bar d d\bar u$ in
    the~SM (without~QCD, CKM mixing and masses in
    unitarity gauge) with one grove of size~31, 
    two of size 12 and two of size 8. Solid lines
    represent gauge flips and dotted lines represent flavor flips.}
\end{figure}
There are 15~five-point tree diagrams in~$\phi^3$ and only four other
diagrams can be reached from any 
diagram by elementary flips.  Thus there is a
non-trivial mathematical structure on the set~$T_5$ with the
relation~$\circ$, visualized by the graph on
the left hand side of
figure~\ref{fig:5phi3/mix71}.

Thus the trivial relation on~$T_4$ has a non-trivial
natural extension to the set~$T_n$ of all $n$-point tree diagrams:
$t\circ t'$ is true if and only if~$t$ and~$t'$ are identical up to a
single flip of a four-point subdiagram
\begin{equation}
   t\circ t' \Longleftrightarrow
     \exists t_4\in T_4,t'_4\in T_4:
       t_4 \circ t'_4 \land t\setminus t_4 = t'\setminus t'_4
\end{equation}
Note that this relation is not transitive and therefore not an
equivalence relation.  Instead, this relation allows us to view the
elements of~$T_n$ as the vertices
of a graph~$F_n$, where the edges of the graph are formed by the pairs
of diagrams related by a single flip
\begin{equation}
\label{eq:forest}
  F_n = \bigl\{(t,t')\in T_n\times T_n \bigl| t\circ t'\bigl\}\,.
\end{equation}
\begin{theorem}
\label{th:vanilla}
  The unflavored forest~$F_n$ is connected for
  all~$n$.\cite{Boos/Ohl:1999:groves,Ohl:1999:habil}
\end{theorem}
Already the simplest non-trivial example of an
unflavored forest, shown on the left
hand side of figure~\ref{fig:5phi3/mix71}, displays an intriguing
symmetry structure: there are 120 permutations of the vertices
of~$F_5$ that leave this forest invariant.

%%%%%%%%%%%%%%%%%%%%%%%%%%%%%%%%%%%%%%%%%%%%%%%%%%%%%%%%%%%%%%%%%%%%%%%%
\section{Groves}
The construction of the groves is based on the observation that the
flips in gauge theories fall into two different classes: the
\emph{flavor flips} among
\begin{equation}
\label{eq:F}
  \{t_4^{F,1},t_4^{F,2},t_4^{F,3}\} = \left\{
   \parbox{16\unitlength}{%
     \begin{fmfgraph}(15,10)
       \fmfleft{f1,f2}
       \fmfright{f1',f2'}
       \fmf{fermion}{f1,v,f2}
       \fmf{fermion}{f2',v',f1'}
       \fmf{boson,tension=0.5}{v,v'}
       \fmfdot{v,v'}
     \end{fmfgraph}},
   \parbox{16\unitlength}{%
     \begin{fmfgraph}(15,10)
       \fmfleft{f1,f2}
       \fmfright{f1',f2'}
       \fmf{fermion}{f1,v1,f1'}
       \fmf{fermion}{f2',v2,f2}
       \fmf{boson,tension=0.5}{v1,v2}
       \fmfdot{v1,v2}
     \end{fmfgraph}},
   \parbox{16\unitlength}{%
     \begin{fmfgraph}(15,10)
       \fmfleft{f1,f2}
       \fmfright{f1',f2'}
       \fmf{phantom}{f1,v1,f1'}
       \fmf{phantom}{f2',v2,f2}
       \fmf{boson,tension=0.5}{v1,v2}
       \fmfdot{v1,v2}
       \fmffreeze
       \fmf{fermion}{f1,v1,f2'}
       \fmf{fermion,rubout}{f1',v2,f2}
     \end{fmfgraph}}
   \right\}\,,
\end{equation}
which involve four matter fields that carry gauge
charge and possibly additional conserved quantum numbers
and the \emph{gauge flips} among
\begin{subequations}
\begin{equation}
\label{eq:G}
  \{t_4^{G,1},t_4^{G,2},t_4^{G,3},t_4^{G,4}\} = \left\{
   \parbox{16\unitlength}{%
     \begin{fmfgraph}(15,10)
       \fmfleft{g1,g2}
       \fmfright{g1',g2'}
       \fmf{boson}{g1,v,g2}
       \fmf{boson}{g1',v',g2'}
       \fmf{boson,tension=0.5}{v,v'}
       \fmfdot{v,v'}
     \end{fmfgraph}},
   \parbox{16\unitlength}{%
     \begin{fmfgraph}(15,10)
       \fmfleft{g1,g2}
       \fmfright{g1',g2'}
       \fmf{boson}{g1,v1,g1'}
       \fmf{boson}{g2,v2,g2'}
       \fmf{boson,tension=0.5}{v1,v2}
       \fmfdot{v1,v2}
     \end{fmfgraph}},
   \parbox{16\unitlength}{%
     \begin{fmfgraph}(15,10)
       \fmfleft{g1,g2}
       \fmfright{g1',g2'}
       \fmf{phantom}{g1,v1,g1'}
       \fmf{phantom}{g2,v2,g2'}
       \fmf{boson,tension=0.5}{v1,v2}
       \fmfdot{v1,v2}
       \fmffreeze
       \fmf{boson}{g1,v1,g2'}
       \fmf{boson,rubout}{g2,v2,g1'}
     \end{fmfgraph}},
   \parbox{16\unitlength}{%
     \begin{fmfgraph}(15,10)
       \fmfleft{g1,g2}
       \fmfright{g1',g2'}
       \fmf{boson}{g1,v,g1'}
       \fmf{boson}{g2,v,g2'}
       \fmfdot{v}
     \end{fmfgraph}}
    \right\}
\end{equation}
or
\begin{equation}
\label{eq:C}
  \{t_4^{G,5},t_4^{G,6},t_4^{G,7},t_4^{G,8}\} = \left\{
   \parbox{16\unitlength}{%
     \begin{fmfgraph}(15,10)
       \fmfright{g,g'}
       \fmfleft{f,f'}
       \fmf{fermion}{f,v}
       \fmf{fermion,tension=0.5}{v,v'}
       \fmf{fermion}{v',f'}
       \fmf{boson}{v,g}
       \fmf{boson}{v',g'}
       \fmfdot{v,v'}
     \end{fmfgraph}},
   \parbox{16\unitlength}{%
     \begin{fmfgraph}(15,10)
       \fmfright{g,g'}
       \fmfleft{f,f'}
       \fmf{fermion}{f,v}
       \fmf{fermion,tension=0.5}{v,v'}
       \fmf{fermion}{v',f'}
       \fmf{phantom}{v,g}
       \fmf{phantom}{v',g'}
       \fmffreeze
       \fmf{boson}{v',g}
       \fmf{boson,rubout}{v,g'}
       \fmfdot{v,v'}
     \end{fmfgraph}},
   \parbox{16\unitlength}{%
     \begin{fmfgraph}(15,10)
       \fmfright{g,g'}
       \fmfleft{f,f'}
       \fmf{fermion}{f,v,f'}
       \fmf{boson,tension=0.5}{v,v'}
       \fmf{boson}{v',g}
       \fmf{boson}{v',g'}
       \fmfdot{v,v'}
     \end{fmfgraph}},
   \parbox{16\unitlength}{%
     \begin{fmfgraph}(15,10)
       \fmfright{g,g'}
       \fmfleft{f,f'}
       \fmf{fermion}{f,v,f'}
       \fmf{boson}{v,g}
       \fmf{boson}{v,g'}
       \fmfdot{v}
     \end{fmfgraph}}
   \right\}\,,
\end{equation}
\end{subequations}
which also involve external gauge bosons
(the diagram $t_4^{G,8}$, is only present for
scalar matter fields).  The flavor flips~(\ref{eq:F}) are special
because they can be switched off without spoiling gauge invariance by
a horizontal symmetry orthogonal to the gauge group.
This suggests to introduce two relations:
\begin{subequations}
\begin{align}
  t\bullet t'  &\Longleftrightarrow
    \text{$t$ and~$t'$ related by a \emph{gauge} flip} \\
  t\circ t' &\Longleftrightarrow
    \text{$t$ and~$t'$ related by a \emph{flavor or gauge} flip}.
\end{align}
\end{subequations}
These define two different graphs on the same set~$T(E)$ of
all Feynman diagrams:
\begin{subequations}
\begin{align}
  F(E) &= \bigl\{(t,t')\in T(E)\times T(E) \bigl| t\circ t'\bigl\} \\
  G(E) &= \bigl\{(t,t')\in T(E)\times T(E) \bigl| t\bullet t'\bigr\}.
\end{align}
\end{subequations}
As we have seen, it is in general not possible to connect all pairs of
diagrams in the \emph{gauge forest}~$G(E)$ by a series of gauge flips
and there is more than one connected component (see
figure~\ref{fig:CC-flips}).  We call the connected components~$G_i(E)$
of~$G(E)$ the \emph{groves} of~$E$.  Since~$t\bullet t'\Rightarrow
t\circ t'$, we have~$\bigcup_i G_i(E)=G(E)\subseteq F(E)$, i.\,e.~the
groves are a \emph{partition} of the gauge forest.
\begin{theorem}
  The forest~$F(E)$ for an external state~$E$
  consisting of gauge and matter fields is connected if the fields
  in~$E$ carry no conserved quantum numbers other than the gauge
  charges. The groves~$G_i(E)$ are the minimal
  gauge invariant classes of Feynman
  diagrams.\cite{Boos/Ohl:1999:groves,Ohl:1999:habil}
\end{theorem}

\begin{figure}
  \begin{center}
    \begin{fmfgraph}(20,15)
      \fmfleft{u,dbar}
      \fmfright{c,sbar}
      \fmftop{d1,d2,g,d3}
      \fmf{phantom}{u,v,dbar}
      \fmf{phantom}{v,v'}
      \fmf{phantom}{sbar,v',c}
      \fmffreeze
      \fmf{dbl_plain_arrow}{v,v'}
      \fmf{fermion}{u,v,vg,dbar}
      \fmf{fermion}{sbar,v',c}
      \fmf{photon,tension=0}{vg,g}
      \fmfdot{v,vg,v'}
    \end{fmfgraph}\quad
    \begin{fmfgraph}(20,15)
      \fmfleft{u,dbar}
      \fmfright{c,sbar}
      \fmfbottom{d1,d2,g,d3}
      \fmf{phantom}{u,v,dbar}
      \fmf{phantom}{v,v'}
      \fmf{phantom}{sbar,v',c}
      \fmffreeze
      \fmf{dbl_plain_arrow}{v,v'}
      \fmf{fermion}{u,vg,v,dbar}
      \fmf{fermion}{sbar,v',c}
      \fmf{photon,tension=0}{vg,g}
      \fmfdot{v,vg,v'}
    \end{fmfgraph}\quad
    \begin{fmfgraph}(20,15)
      \fmfleft{u,dbar}
      \fmfright{c,sbar}
      \fmftop{d1,d2,g,d3}
      \fmf{phantom}{u,v,dbar}
      \fmf{phantom}{v,v'}
      \fmf{phantom}{sbar,v',c}
      \fmffreeze
      \fmf{dbl_plain_arrow}{v,vg,v'}
      \fmf{fermion}{u,v,dbar}
      \fmf{fermion}{sbar,v',c}
      \fmf{photon,tension=0}{vg,g}
      \fmfdot{v,vg,v'}
    \end{fmfgraph}\quad
    \begin{fmfgraph}(20,15)
      \fmfleft{u,dbar}
      \fmfright{c,g,sbar}
      \fmf{phantom}{u,v,dbar}
      \fmf{phantom}{v,v'}
      \fmf{phantom}{sbar,v',c}
      \fmffreeze
      \fmf{dbl_plain_arrow}{v,v'}
      \fmf{fermion}{u,v,dbar}
      \fmf{fermion}{sbar,vg,v',c}
      \fmf{photon,tension=0}{vg,g}
      \fmfdot{v,vg,v'}
    \end{fmfgraph}\quad
    \begin{fmfgraph}(20,15)
      \fmfleft{u,dbar}
      \fmfright{c,g,sbar}
      \fmf{phantom}{u,v,dbar}
      \fmf{phantom}{v,v'}
      \fmf{phantom}{sbar,v',c}
      \fmffreeze
      \fmf{dbl_plain_arrow}{v,v'}
      \fmf{fermion}{u,v,dbar}
      \fmf{fermion}{sbar,v',vg,c}
      \fmf{photon,tension=0}{vg,g}
      \fmfdot{v,vg,v'}
    \end{fmfgraph}
  \end{center}
  \caption{\label{fig:CC-flips}%
    All five diagrams in $u\bar d\to c\bar s\gamma$ are in the same
    grove, because they are connected by gauge flips passing through
    the diagram in the center.  In contrast,
    in~$e^+e^-\to\mu^+\mu^-\gamma$, the center diagram is
    missing and there are two separate groves.}
\end{figure}
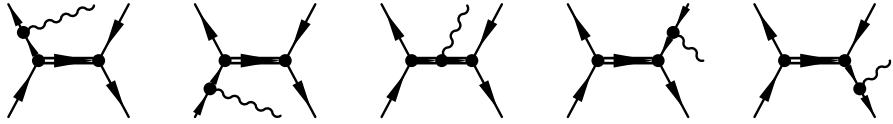
The forests and groves are very symmetrical and their automorphism
groups turn out to be surprisingly large: the automorphism group of the
forest~$F(\gamma\gamma\to u\bar d d\bar u)$ on the right hand side of
figure~\ref{fig:5phi3/mix71} has 128~elements.

%%%%%%%%%%%%%%%%%%%%%%%%%%%%%%%%%%%%%%%%%%%%%%%%%%%%%%%%%%%%%%%%%%%%%%%%
\section*{Acknowledgments}
This work was supported in part by DFG (MA\,676/5-1) and RFBR (99-02-04011).
E.\,B.~is grateful to the Russian Ministry of Science and
Technologies, and to the Sankt-Petersburg Grant Center
for partial financial support. T.\,O.~is supported by BMBF, Germany
(05\,7SI79P\,6, 05\,HT9RDA).
%%%%%%%%%%%%%%%%%%%%%%%%%%%%%%%%%%%%%%%%%%%%%%%%%%%%%%%%%%%%%%%%%%%%%%%%

%%%%%%%%%%%%%%%%%%%%%%%%%%%%%%%%%%%%%%%%%%%%%%%%%%%%%%%%%%%%%%%%%%%%%%%%
\end{empfile}
\end{fmffile}
\end{document}